\documentclass[aps,prd,reprint,onecolumn,nofootinbib,notitlepage]{revtex4-1}

\usepackage[T1]{fontenc} 

\usepackage{graphicx}
\usepackage{subfig}
\usepackage{hyperref}
\usepackage{amsmath}
\usepackage{color}
\usepackage{amssymb}
\usepackage[english]{babel}
\usepackage{epsfig}

\usepackage[utf8]{inputenc}
\usepackage[english]{babel}
\usepackage{amsfonts}
\usepackage{amssymb}

\begin{document}

\title{\boldmath Cosmological constraints on a unified dark matter-energy scalar field model\\ with fast transition}

\author{Iker Leanizbarrutia$^{1}$}
\email{iker.leanizbarrutia@ehu.eus}
\author{Alberto Rozas-Fern\'andez$^{2}$}
\email{a.rozas@oal.ul.pt}
\author{Ismael Tereno$^{2,3}$}
\email{tereno@fc.ul.pt}

\affiliation{
	${}^1$ Department of Theoretical Physics, University of the Basque Country UPV/EHU, P.O. Box 644, 48080 Bilbao, Spain\\
	${}^2$ Instituto de Astrof\'{\i}sica e Ci\^{e}ncias do Espa\c{c}o, Universidade de Lisboa, OAL, Tapada da Ajuda, PT1349-018 Lisboa, Portugal\\
	${}^3$ Departamento de F\'{\i}sica, Faculdade de Ci\^encias, Universidade de Lisboa, Edif\'{\i}cio C8, Campo Grande, PT1749-016 Lisbon, Portugal}

\begin{abstract}
	We test the viability of a single fluid cosmological model containing a transition from a dark-matter-like regime to a dark-energy-like regime. The fluid is a k-essence scalar field with a well-defined Lagrangian. We constrain its model parameters with a combination of geometric probes and conclude that the evidence for this model is similar to the evidence for $\Lambda$CDM.  In addition, we find a lower bound for the rapidity of the transition, implying that fast transitions are favored with respect to slow ones even at background level.
\end{abstract}

\maketitle

\section{Introduction}
\label{sec:intro}

In the past two decades, a variety of cosmological data \cite{DWeinberg:2013} has been pointing to the conclusion that the expansion of the Universe is accelerating at present.
The favored explanation, the $\Lambda$CDM model, constitutes the standard cosmological paradigm. In this model, $\Lambda$ is the cosmological constant, which drives the accelerated expansion, and cold dark matter (CDM) forms the large-scale structures in the Universe. However, the model suffers from theoretical and conceptual issues, such as the cosmological constant and coincidence problems, as well as observational challenges, such as the description of small-scale cosmological structures, see \cite{Bull:2016} for a review. For these reasons, a long list of alternatives have been explored, mainly in the form of dynamical dark energy (DE) or modifications of general relativity (see \cite{Copeland:2006wr,Clifton:2011jh} for reviews). Very different alternatives, also compliant with some of the cosmological data, have been proposed such as models where the acceleration effects are explained by quantum effects \cite{GonzalezDiaz:2006tr,GonzalezDiaz:2008ci,GonzalezDiaz:2008ba,RozasFernandez2017}, or also the averaging approach to cosmology \cite {Smale:2011}.

Here we will consider a well-studied variation of $\Lambda$CDM: the unified dark fluid approach, also known as quartessence, sometimes as unified dark energy, but more usually known as unified dark matter (UDM)\footnote{The name ``unified dark matter'' that prevails in the literature is misleading. To our knowledge, it is a colloquial simplification of  ``unified dark matter-energy'', which was the original meaning of the acronym UDM first proposed in \cite{Makler:2002jv}. We will follow the original proposal, reinstating the naming ``unified dark matter-energy'' as the meaning of the well-established acronym UDM.}. A plethora of UDM models have been proposed (see \cite{Bertacca:2010ct} for a review) after the pioneering introduction of the Chaplygin gas \cite{Kamenshchik:2001cp,Bilic:2001cg,Bento:2002ps}.
The unification of dark matter and dark energy is an interesting approach that assumes the existence of a single fluid capable of accounting for both the accelerated expansion at late times and the large-scale structure formation at early times, due to the evolution of its equation of state (EOS) and speed of sound. In principle this is more efficient than postulating two different fluids and equally valid, since the nature of the fluids is still elusive. It also has the advantage of evading, by definition, the coincidence problem \cite{cp}. These models substantially alleviate as well the tension between some recent high and low redshift measurements \cite{Camera:2017tws}.

A serious issue in most UDM models is the presence of an effective speed of sound that can be very different from zero during the cosmological evolution. This prevents the dark fluid to cluster below a thresholding scale (the Jeans scale) \cite{Hu:1998kj,Garriga:1999vw,Pietrobon:2008js}. In addition, the evolution of the gravitational potential may also give rise to a strong signature in the integrated Sachs Wolfe (ISW) effect \cite{Bertacca:2007cv}. It is therefore crucial to make sure that the single dark fluid is able to cluster and create the observed cosmic structures as well as reproducing the well-known pattern of cosmic microwave background (CMB) temperature anisotropies \cite{Carturan:2002si}. However, for the majority of UDM models in the literature, these requirements, together with the necessity of having a background evolution that complies with observations, lead to a severe fine-tuning of the parameters, to the point that the models become almost indistinguishable from $\Lambda$CDM and are thus less interesting \cite{Sandvik:2002jz, Scherrer:2004au, Giannakis:2005kr, Piattella:2009da}.

The problem of the lack of clustering, or production of oscillations, can be avoided with a technique introduced in \cite{Bertacca:2008uf}. In particular, the dark fluid is a scalar field, $\varphi$, with a noncanonical kinetic term, i.e., a term $f(\dot\varphi^2)$ instead of the standard $\dot\varphi^2/2$. In this way it was possible to build a UDM model with a small effective sound speed that allows structure formation and has a weak ISW effect, being compliant with weak lensing data \cite{Camera:2009uz, Camera:2010wm}. This model has, however, the same background evolution as $\Lambda$CDM. A more recent alternative are the so-called UDM models with fast transition where, during a short period, the effective speed of sound can be large, but is otherwise zero. This produces a fast transition between a CDM-like era, with an  Einstein-de Sitter evolution, and an accelerated DE-like era, and allows for structure formation. In addition, these models are not forced by construction to have the same background evolution as $\Lambda$CDM and are free from the problem of fine-tuning of the parameters that plagues many UDM models. The thermodynamics of a UDM model with fast transition was explored in \cite{Radicella:2014nka}.

The dynamics of UDM models with fast transition can be prescribed in three different ways: starting from either the EOS $w$, the pressure $p$ or the energy density $\rho$. The first UDM model with fast transition was introduced in \cite{Piattella:2009kt} and prescribed the evolution of $p$. The pressure and energy density were related by a barotropic EOS, $p=p(\rho)$ and the perturbations were adiabatic. A second UDM model with fast transition was presented in \cite{Bertacca:2010mt} and was built from a k-essence \cite{Chiba:1999ka, ArmendarizPicon:2000ah} scalar field Lagrangian (see also \cite{Kang:2007vs, Cruz:2008cwa, RozasFernandez:2011je, Rozas-Fernandez:2014tsa}). This model also prescribed $p$ but, differently from the first one, since it is based on a scalar field the perturbations are naturally nonadiabatic \cite{DiezTejedor:2005fz, Bilic:2008zk}, allowing for a small Jeans length even when the speed of sound is non-negligible. The model also contains a future attractor that acts as an effective cosmological constant\footnote{A scalar field with a potential that admits a minimum $V_{0}=V(\phi_{0})\not =0$ is equivalent to a cosmological constant $\rho_{\Lambda}=V_{0}$ and a scalar field in a potential $\tilde{V}=V-V_{0}$.}, $\rho_{\infty}$; i.e., an asymptotic limit $w = -1$ is built in. A third UDM model with fast transition was proposed in \cite{Bruni:2012sn}. This is a phenomenological model, with the dynamics prescribed through the fluid density $\rho$, and it has adiabatic perturbations.

Models with a fast transition might also be a step towards a unified description of dark matter, dark energy and inflation \cite{Liddle:2006qz} but, regardless of that possibility, they are considered among the most promising UDM models \cite{Amendola:2016saw}. Even though they are built with the goal of enabling structure formation, it is also important to test them at the background level since they may have a background evolution quite distinct from $\Lambda$CDM. In particular, such tests will constrain the rapidity of the transition and may already give an indication whether the allowed rapidity range favors structure formation. The phenomenological UDM model, and variations of it, were recently constrained at background level in \cite{Lazkoz:2016hmh}. In the present work, we apply supernova, galaxy clustering and CMB data to test the scalar field UDM model of \cite{Bertacca:2010mt}, constraining its parameters and making a statistical model comparison with both $\Lambda$CDM and the phenomenological UDM model of \cite{Bruni:2012sn} tested in \cite{Lazkoz:2016hmh}.

In the rest of the paper, we present in Sec.~\ref{sec:UDMmodelfast} the UDM model that will be tested in Sec.~\ref{sec:results} using the data and methods described in Sec.~\ref{sec:constraints}. We conclude with a summary and some remarks in Sec.~\ref{sec:concl}.

\section{The UDM model}
\label{sec:UDMmodelfast}

We consider the scalar field UDM model proposed in \cite{Bertacca:2010mt}, where the evolution of the pressure has the following form:
\begin{equation}
\label{ptanh}
p(a) = -\rho_\infty \left\{\frac{1}{2} + \frac{1}{2}\tanh\left[\frac{\beta}{3} \left(a^3 - a_{\rm t}^3\right)\right]\right\}\,.
\end{equation}
This model allows for a fast transition in the pressure evolution, since for large values of $\beta$ the $\tanh$ function tends to a step function. The transition occurs at a scale factor $a_t$, with rapidity parameterized by $\beta$, while $\rho_\infty$ parameterizes the pressure amplitude. The fluid goes from an Einstein-de Sitter DM era ($p=0$ at early times), through $p(a_t)=-\rho_\infty/2$ at transition, to a DE era at late times (with $p$ reaching $-\rho_\infty$ the sooner for faster transitions).

Considering a Friedmann-Lema\^{i}tre-Robertson-Walker (FLRW) background metric (and a frame with proper time coinciding with the cosmic time), the density can be derived from the pressure using the energy conservation equation
\begin{equation}
\label{coneq}
 \dot{\rho} = -3H\left(\rho + p\right)=-3H\rho\left(1 + w\right)\,,
\end{equation} 
where $w = p/\rho$ is the EOS and the dot means differentiation with respect to time. The density is obtained from the pressure by integrating Eq.~(\ref{coneq}):
\begin{equation}
\label{tanh}
\rho(a) = \rho_\infty \left\{\frac{1}{2}+\frac{3}{2\beta}a^{-3}\ln\left(\cosh\left[\frac{\beta}{3} \left(a^3 - a_{\rm t}^3\right)\right]\right)\right\} + \rho_{c0} a^{-3}\,.
\end{equation}
The integration introduces another constant. It is usual to choose it as the amplitude of a "CDM sector of the UDM": $\rho_{c0}$, defined at $a=1$. Note that the density does not have a fast transition, since the $\ln(\cosh)$ function is not a step function. The density decreases smoothly from its maximum amplitude at $a=0$, through $\rho(a_t) =(\rho_{c0}a_t^{-3}+\rho_\infty)$ at transition, to $\rho_\infty$ when $a \rightarrow \infty$. Note also that for fast transitions (large $\beta$) and after the transition, $\tanh \sim 1$ and $\ln[\cosh(x)]/x \sim 1$, and thus $p \sim -\rho_\infty$ and $w \sim -\rho_\infty / (\rho_\infty + \rho_{c0})$. This means that fastest models become degenerate and are more similar to $\Lambda$CDM than the slower ones (with the exception of the singular case $\beta=0$).

The UDM model contain thus four parameters: $\rho_{c0}$, $\rho_\infty$, $\beta$ and $a_t$. With this choice of parameters, the density is written as the sum of three parts: the CDM-like term $\rho_c(a)=\rho_{c0} a^{-3}$, a constant term $\rho(a)=\rho_\infty / 2$ and the $\ln\cosh$ term, with the latter two defining a ``dark energy sector''. To compare UDM models with $\Lambda$CDM, it is useful to define today's densities for these two sectors. Introducing the critical density today, $\rho_{\rm cr}=3H_0^{2}$, we define the two dimensionless density parameters:
\begin{equation}
\Omega_c = \frac{\rho_{c0}}{3H_0^2}
\end{equation}
and
\begin{equation}
\label{omegaDEdef}
\Omega_{DE}=\frac{\rho_\infty}{3H_0^2} \left\{\frac{1}{2}+\frac{3}{2\beta}\ln\left(\cosh\left[\frac{\beta}{3} \left(1 - a_{\rm t}^3\right)\right]  \right) \right\}\,.
\end{equation}

All the background probes we will use in the likelihood analysis depend on the Hubble function
\begin{equation}\label{friedman-2}
E^2(a)=H^2/H_0^2=\Omega_{r} a^{-4}+ \Omega_{b} a^{-3}+ \Omega_{\rm UDM}(a)\,,
\end{equation}
where $\Omega_b$ and $\Omega_r=2.49\times10^{-5}h^{-2}$ are the baryonic matter and radiation densities, respectively, and
\begin{equation}\label{omegaUDM}
\Omega_{\rm UDM}(a)=\Omega_{c}a^{-3}+\Omega_{DE}\left\{ \frac{1}{2}+\frac{3}{2\beta}a^{-3}\ln\left\{\cosh\left[\frac{\beta}{3} \left(a^3 - a_{\rm t}^3\right)\right]  \right\} \right\}\;.
\end{equation}
The four parameters $\Omega_{DE}$, $\Omega_c$, $\beta$ and $a_t$ are not all independent. Indeed, applying Friedmann's equation, $\sum_i\Omega_i =1$, we can write
\begin{equation}
\label{omegaDE}
\Omega_{DE}=\frac{1-\Omega_{r}-\Omega_{b}-\Omega_{c}}{ \frac{1}{2}+\frac{3}{2\beta}\ln\left\{\cosh\left[\frac{\beta}{3} \left(1 - a_{\rm t}^3\right)\right]  \right\} }.
\end{equation}

The definition of two sectors allows the introduction of an EOS of the dark energy sector,
\begin{equation}
\label{wDE}
w_{\rm DE}(a)=\frac{p(a)}{\rho(a)-\rho_{\rm c0}a^{-3}}\;,
\end{equation}
in addition to the EOS $w(a)=p(a)/\rho(a)$.

We finally note that an explicit analytical Lagrangian can be written for this model, since the general Lagrangian for a UDM scalar field $\varphi$, within the framework of k-essence, is 
\begin{equation}
L=  L_G+ L_\varphi=\frac{1}{16\pi G}R+ L_\varphi(\varphi,X),
\end{equation}
where $X$ is the kinetic term and the pressure can be identified with the term $p=L_\varphi(\varphi,X)$.

\section{Methodology}
\label{sec:constraints}

We test the model with a Markov chain Monte Carlo (MCMC) exploration of the parameter space \cite{Christensen:2001gj,Lewis:2002ah}, combining various probes of the expansion history of the Universe: luminosity distances to type Ia supernovae, baryon acoustic oscillation scale parameter, Alcock-Paczynski distortion parameter, and include CMB distance priors. The various data sets are uncorrelated and thus the total $\chi^2$ used in the analysis is simply the sum
\begin{equation}
\chi^2 = \chi^{2}_{CMB} + \chi^{2}_{BAO} + \chi^{2}_{SN}\;.
\end{equation}

\subsection{SNe Ia data}

As in the previous analysis \cite{Lazkoz:2016hmh}, we use the Union2.1 compilation  \cite{Suzuki:2011hu}, which provides not only the distance modulus $\mu(z_i)$ for each SN, but also the full statistical plus systematics covariance matrix. The data set consists of $580$ type Ia supernovae with redshifts in the interval $0.015 < z < 1.414$. The cosmological model is tested through the dimensionless luminosity distance
\begin{equation}
d_L(z) = (1+z) \int_0^z \frac{dz'}{E(z')}\, ,
\end{equation}
which depends on the dimensionless Hubble function $E(z)$ and is directly related to the observable: the distance modulus
\begin{equation}
\mu(z) = 5 \log_{10} d_L(z) + \mu_{0} \;.
\end{equation}
This relation includes an additive nuisance parameter, $\mu_{0}$, involving the values of the speed of light $c$, Hubble constant $H_{0}$, and SNe Ia absolute magnitudes.
The likelihood is assumed to be Gaussian and is defined as
\begin{equation}
\chi^2_{SN}=\textbf{X}_{SN}^T \cdot \textbf{C}_{SN}^{-1} \cdot \textbf{X}_{SN} \,,
\end{equation}
where $\textbf{X}_{SN}$ is the difference vector with elements $X_{iSN}=\mu_{\rm model}(z_{i}) - \mu_{\rm obs}(z_{i})$ and $\textbf{C}_{SN}$ is the data covariance matrix. We analytically marginalize over the additive parameter $\mu_0$, as an alternative to including it in the MCMC parameter space. The resulting $\chi^2$ is given by \cite{Conley:2011ku}
\begin{equation}
\chi^2_{SN} = a + \log \frac{d}{2 \pi } - \frac{b^2}{d} \, ;
\end{equation}
where $a \equiv  \textbf{X}_{SN}^{T} \cdot \textbf{C}_{SN}^{-1} \cdot  \textbf{X}_{SN}$, $b \equiv  \textbf{X}_{SN}^{T} \cdot \textbf{C}_{SN} ^{-1} \cdot \bf{1}$, and $d \equiv {\bf 1}^{T} \cdot \textbf{C}_{SN}^{-1} \cdot \bf{1}$, with $\bf{1}$ being the identity matrix.

\subsection{Baryon acoustic oscillation (BAO) data}

Unlike the previous analysis \cite{Lazkoz:2016hmh},  we will now use the baryon acoustic oscillation scale parameter $A(z)$ and the Alcock-Paczynski distortion parameter $F(z)$ provided by the WiggleZ Dark Energy Survey \cite{Blake:2012pj}, as the BAO observables. They are defined as
\begin{eqnarray}
A(z) & \equiv & 100 D_V(z) \sqrt{\Omega_m h^2} /cz \, , \\
F (z) & \equiv & (1+z) D_A(z) H(z)/c \, ,
\end{eqnarray}
probing the angular-diameter distance
\begin{equation}
D_A(z) = \frac{c}{(1+z)} \int^z_0 \frac{dz'}{H(z')}
\end{equation}
and the volume-averaged distance
\begin{equation}
D_V (z) = \left[ (1+z)^2 D_A(z)^2 \frac{c z}{H(z)} \right]^{1/3} \, .
\end{equation}
WiggleZ measured these observables in three overlapping redshift bins, with effective redshifts $(z_{1}, \, z_{2}, \, z_{3}) = (0.44, \, 0.60, \, 0.73)$. The data values are
\begin{eqnarray}
\textbf{X}_{obs} &=& (A_1, A_2, A_3, F_1, F_2, F_3) \\
&=& (0.474, 0.442, 0.424, 0.482, 0.650, 0.865)
\end{eqnarray}
with correlated errors described by the covariance matrix
\begin{eqnarray}
\textbf{C}_{BAO}=10^{-3} \times
\left( \begin{array}{cccccc}
	1.156 & 0.211 & 0.0   & 0.400 & 0.234 & 0.0 \\
	0.211 & 0.400 & 0.189 & 0.118 & 0.276 & 0.336 \\
	0.0   & 0.189 & 0.441 & 0.0   & 0.167 & 0.399 \\
	0.400 & 0.118 & 0.0   & 2.401 & 1.350 & 0.0 \\
	0.234 & 0.276 & 0.167 & 1.350 & 2.809 & 1.934 \\
	0.0   & 0.336 & 0.399 & 0.0   & 1.934 & 5.329 \\
\end{array} \right)
\end{eqnarray}
The BAO contribution to the total $\chi^2$ is
\begin{equation}
\chi^2_{BAO} = \textbf{X}_{BAO}^T \cdot \textbf{C}_{BAO}^{-1} \cdot \textbf{X}_{BAO}
\end{equation}
where $\textbf{X}_{BAO} = (\textbf{X}_{obs} - \textbf{X}_{mod} )$ is the difference vector.

\subsection{Priors}

In order to reduce the volume of the parameter space in the MCMC analysis, it is useful to include the so-called distance priors \cite{Wang2013} in our analysis. These are priors on the CMB shift parameters, geometrical quantities that effectively summarize the CMB data, since they capture the degeneracies between the parameters that determine the CMB power spectrum  \cite{WangMukherjee2007}.

The first shift parameter is a dimensionless distance to the photon-decoupling surface
\begin{equation}
R \equiv \sqrt{\Omega_m H^2_0} \,\frac{r(z_*)}{c} \;,
\end{equation}
defined from the comoving distance to the photon-decoupling surface
\begin{equation}
r(z_*) = c\,\int_0^{z_*} \frac{dz'}{H(z')}\,.
\end{equation}
The redshift of the photon-decoupling surface may be computed from a fitting formula
\cite{Hu:1995en}
\begin{equation}
z_{*}= 1048 \left[ 1+0.00124(\Omega_b h^2)^{-0.738} \right] \left[ 1 + g_1 (\Omega_m h^2)^{g_2} \right] ,
\end{equation}
where $g_1$ and $g_2$ are functions of the physical baryon density.

The second shift parameter is a dimensionless size of the sound horizon at the photon-decoupling epoch, i.e., the angular scale of the sound horizon at the photon-decoupling epoch
\begin{equation}
l_a \equiv \pi \frac{r(z_*)}{r_s(z_*)} \;,
\end{equation}
defined from the comoving sound horizon
\begin{eqnarray}
r_s(z_*) &=& \int_0^{a_*} \frac{da'}{a'^2} \frac{c_s}{H(a')}  ,
\end{eqnarray}
where the sound speed $c_s$ depends on the physical baryon density and the temperature of the CMB \cite{Wang2013}. We take $T_{CMB}=2.725$ K \cite{Fixsen:2009ug}.

The distance prior we use is the Gaussian fit to the joint probability density function of $R$ and $l_a$ presented in \cite{Wang2013} and derived from \textit{Planck} first release data \cite{Ade:2013zuv} and \textit{WMAP} 7 \cite{wmap7} and \textit{WMAP9} \cite{wmap9} temperature and polarization data. Since the shift parameters correlate with the physical baryon density $\Omega_b h^2$ the prior also includes the baryon density and it is a three-dimensional Gaussian with mean
\begin{equation}
\left(\langle l_a \rangle , \langle R \rangle, \langle \Omega_b h^2 \rangle\right) =
(301.57 , 1.7407 , 0.02228)\,
\end{equation}
and covariance
\begin{eqnarray}\label{Vcmb}
\textbf{C}_{CMB} = 10^{-2} \times
\left( \begin{array}{ccc}
	3.24 & 0.08883 & -0.0022869\\
	0.08883 & 0.008836 & -1.953\times 10^{-4}\\
	-0.0022869 & -1.953 \times 10^{-4} & 9.\times 10^{-6}\\
\end{array} \right) \; .
\end{eqnarray}
The CMB data contribution to the total $\chi^2$ is thus
\begin{equation}
\chi^2_{CMB} = \textbf{X}_{CMB}^T \cdot \textbf{C}_{CMB}^{-1} \cdot \textbf{X}_{CMB} \; ,
\end{equation}
where the three-dimensional difference vector between model and observations is
\begin{eqnarray} \label{data-vector}
\textbf{X}_{CMB} = \left( \begin{array}{c}
	l_a - \langle l_a \rangle \\
	R - \langle R \rangle \\
	\Omega_b h^2 - \langle \Omega_b h^2 \rangle \end{array} \right) \; .
\end{eqnarray}

Besides the distance priors, we also include some broad and flat conditions: the dark matter density must be positive $0<\Omega_{c}<1$; the baryonic matter density must be positive and smaller than the dark matter density $0 < \Omega_b < \Omega_{c}$; the Hubble function must be positive for all values of the scale factor $a$, $ E(a) > 0 $ ; and $0 < a_t < 1 $ because we want the transition to actually have happened. Finally, the Hubble constant $H_0$ is analytically marginalized in the SN likelihood and is left as a free parameter, with a broad flat prior, in the BAO and CMB likelihoods.

\section{Analysis and Results}
\label{sec:results}

We ran a set of Markov chains on the five-dimensional parameter space $(h, \Omega_c, \Omega_b, a_t, \beta)$, using the following three-step procedure. We start by running a short preliminary chain of around 20 000 iterations in order to find the region of maximum probability density. Then we make a second run for around 50 000 iterations to find a tentative covariance matrix. Finally, we start the final chain, using the previously found covariance matrix as a proposal step. The final chain has around 200 000 points and we assess its convergence using the ratio of variances proposed in \cite{Dunkley:2005}. Differently from the more standard Gelman and Rubin ratio of variances that compare parallel chains \cite{Tereno:2005}, the convergence ratio of \cite{Dunkley:2005} uses only one chain. It is based on a spectral analysis of the single MCMC chain and in order to perform the test we compute the power spectrum of the chain on 1000 Fourier modes.

\begin{figure*}
\begin{minipage}{1.0\textwidth}
		\centering
		\includegraphics[width=1.0\textwidth]{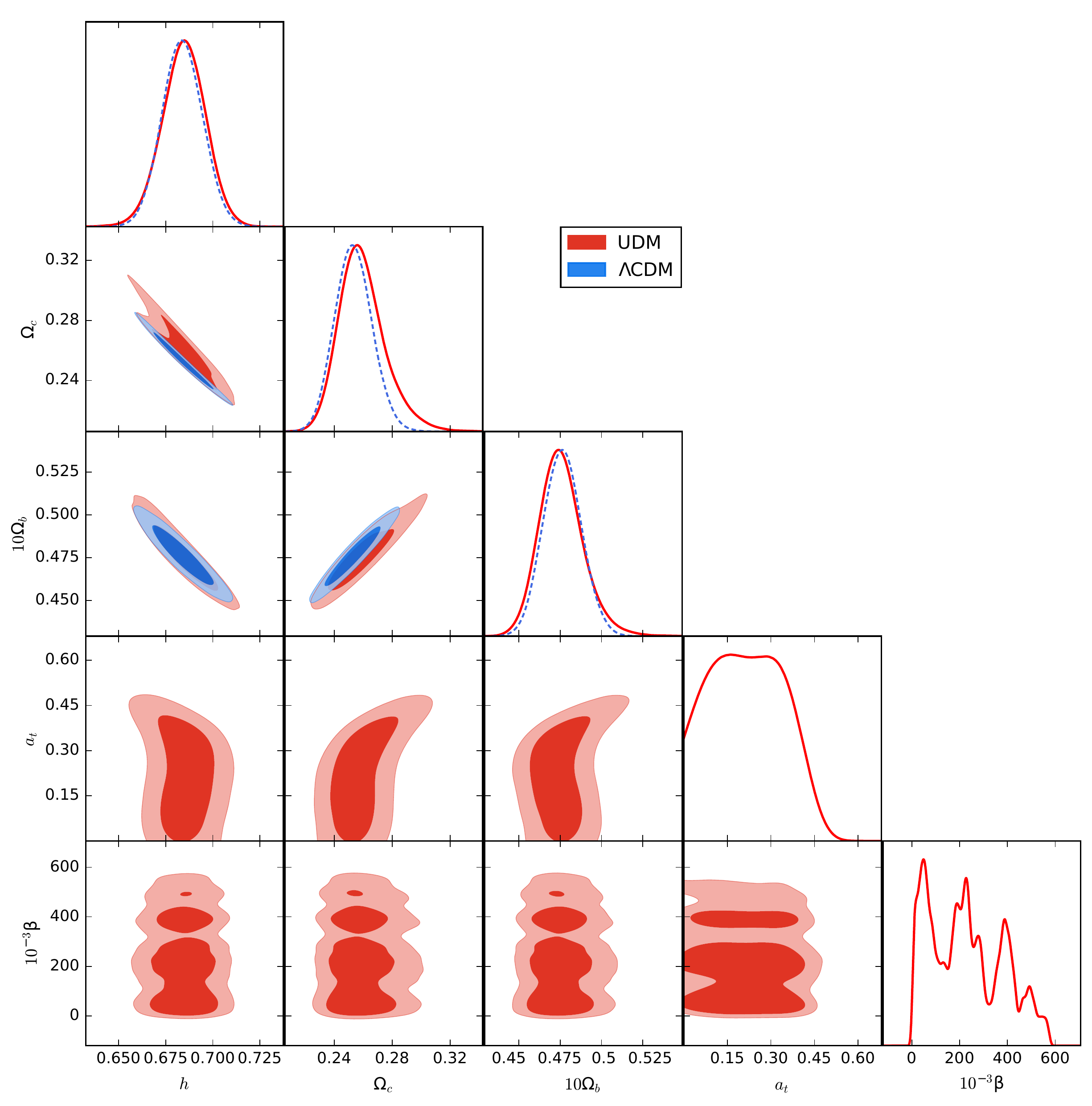}~~~~~
		\caption{Posterior distribution from the full-range MCMC chains.
Diagonal panels: One-dimensional marginalized posterior distributions for UDM (red, solid lines) and $\Lambda$CDM (blue, dashed lines) parameters. Off-diagonal panels: 1- and 2-$\sigma$ two-dimensional marginalized contours for UDM (red) and $\Lambda$CDM (blue) parameters.}
 \label{fig:UDM}
\end{minipage}
\end{figure*}

Besides the UDM scenario, we also ran an MCMC for the $\Lambda$CDM scenario. Figure~\ref{fig:UDM} shows the posterior probabilities for each parameter of the UDM and $\Lambda$CDM models, along with 1- and 2-$\sigma$ two-dimensional confidence regions. Table~\ref{table-UDM} gives the corresponding median and marginalized 1-$\sigma$ interval for each chain parameter and some derived parameters.

\begin{table}[h]
	\centering
	\caption{Median and 1-$\sigma$ uncertainty for the UDM and $\Lambda$CDM model parameters, from the full-range MCMC chains.
	}\label{table-UDM}
	\vspace{5pt}
	\begin{tabular} {lcc}
		\hline
		Parameter &   UDM & $\Lambda$CDM\\
		\hline
		\hline
		$h$ & $0.685\pm 0.012$  & $0.684\pm 0.011$ \\
		$\Omega_c$ & $0.259^{+0.012}_{-0.018}$ & $0.253\pm 0.013$
		\vspace{3pt}
		\\
		$\Omega_b$ & $0.0476^{+0.0011}_{-0.0014}$ & $0.0476\pm 0.0011$
		\vspace{3pt}
		\\
		$a_t$ & $0.22^{+0.13}_{-0.15}$ &
		\vspace{3pt}
		\\
		$\beta$ & $227500^{+200000}_{-200000}$ &
		\vspace{3pt}
		\\		
		\hline
		$\Omega_{DE}               $ & $0.693^{+0.019}_{-0.013}   $&
		\vspace{3pt}
		\\
		$w_{DE}                    $ & $-1.011^{+0.011}_{-0.0039} $&
		\vspace{3pt}
		\\
		\hline
		$w                         $ & $-0.735^{+0.013}_{-0.015}  $&\\
		
		\hline
	\end{tabular}
\end{table}

The constraints on the three standard parameters are similar in the two models. The probability contours of the Hubble parameter vs densities show the usual anti-correlations that arise because Hubble function and distance measurements probe physical densities $\Omega_ih^2$. The main new feature is a slight correlation between the scale factor of transition $a_t$ and $\Omega_c$, especially for higher values of $a_t$ (and a corresponding anti-correlation with $h$). This degeneracy broadens the $\Omega_c$ contours, being responsible for the decrease of precision in the $\Omega_c$ estimate quoted in Table~\ref{table-UDM}. This differs from the behavior found in the analysis of the phenomenological UDM models \cite{Lazkoz:2016hmh}, where the constraint on $\Omega_c$ was found to be stronger than in the $\Lambda$CDM model, even though the evidence was not conclusive in favor of that UDM model.

For model comparison purposes, we start by noticing in Table~\ref{table-comp} that the UDM best fit has a lower $\chi^2$ value than the one found in the $\Lambda$CDM analysis. This may be due to overfitting, and indeed the best-fit reduced $\chi^2$ is larger than for the $\Lambda$CDM case. A more robust way to compare the models is through the ratio of the model evidences, i.e., the Bayes factor \cite{Trotta:2005ar}. We compute the evidence with an implementation of the nested sampling algorithm of \cite{Mukherjee:2005wg}. In particular, we use $10^3$ sample points, chosen randomly, and compute the evidence in up to $10^4$ steps. We repeat the procedure 100 times, varying the sample points, and quote the average evidence from the 100 realizations. We obtain a Bayes factor very close to 0, and thus the model comparison is highly inconclusive, according to  Jeffreys' scale \cite{Gordon:2007xm}.

\begin{table}[h!]
	\centering
	\caption{Values from five methods to perform model comparison between UDM, $\Lambda$CDM and the phenomenological UDM.}\label{table-comp}
	\vspace{5pt}
	\begin{tabular} {cccc}
		\hline
		 & UDM & $\Lambda$CDM & $\rm{UDM_{ph}}$\\
		\hline
		 {$\chi^2_{\rm min}$} & $552.59$& $552.77$  & 552.75\\
		
		 {$\chi^2_{\rm red}$}& $0.9478$ & $0.9449$ & 0.9481 \\
		
		{$\rm \ln B_{U\Lambda}$} & $-0.0196$ & $0$ & 0.6850\\
		
		BIC & 584.485 & 571.902 & 584.644\\
		DIC & 553.250 & 552.770 & 552.814\\

		\hline
	\end{tabular}
\end{table}

With a Bayes factor so close to 0, we decided to investigate if the behavior would be any different when using approximate evidence measures, namely information criteria. The Bayesian information criterion (BIC) is defined as \cite{Liddle:2004}
\begin{equation}
{\rm BIC} = -2\ln{\rm L_{max}} + k \ln{N}.
\end{equation}
Since the number of data points used, $N$, was the same for the two models, BIC directly penalizes the lower minimum $\chi^2$ of UDM with the higher number of free parameters $k$. For the deviance information criterion (DIC), we followed \cite{SaezGomez:2016} and computed
\begin{equation}
{\rm DIC} = 2\left<\chi^2\right> - \chi^2_{\rm \min},
\end{equation}
where the average $\chi^2$ were computed from the chains and not with the nested sampling code. The results are consistent with the comparison of evidence in that both information criteria assign a weak but inconclusive preference to $\Lambda$CDM.

We are also interested in comparing the stronger motivated scalar field UDM model with the phenomenological one. For that purpose we made a new analysis of the latter, testing it with the same set of data used in our present analysis. The results from this second model comparison analysis are also summarized in Table~\ref{table-comp}. Model comparison between the two UDM models is more direct, since both have the same number of parameters and data points. Therefore, BIC reduces to a measure of the best fit, which is slightly in favor of the scalar field model. It is interesting to note that even though the scalar field model shows a better best fit $\chi^2$, it has a worse $\chi^2$ behavior on average and consequently a lower DIC value and evidence. Again, the analysis does not favor one model over the other, with a weak but inconclusive preference for the phenomenological model.

\begin{figure*}[h!]
	\begin{minipage}{1.0\textwidth}
		\centering
		\includegraphics[width=0.5\textwidth]{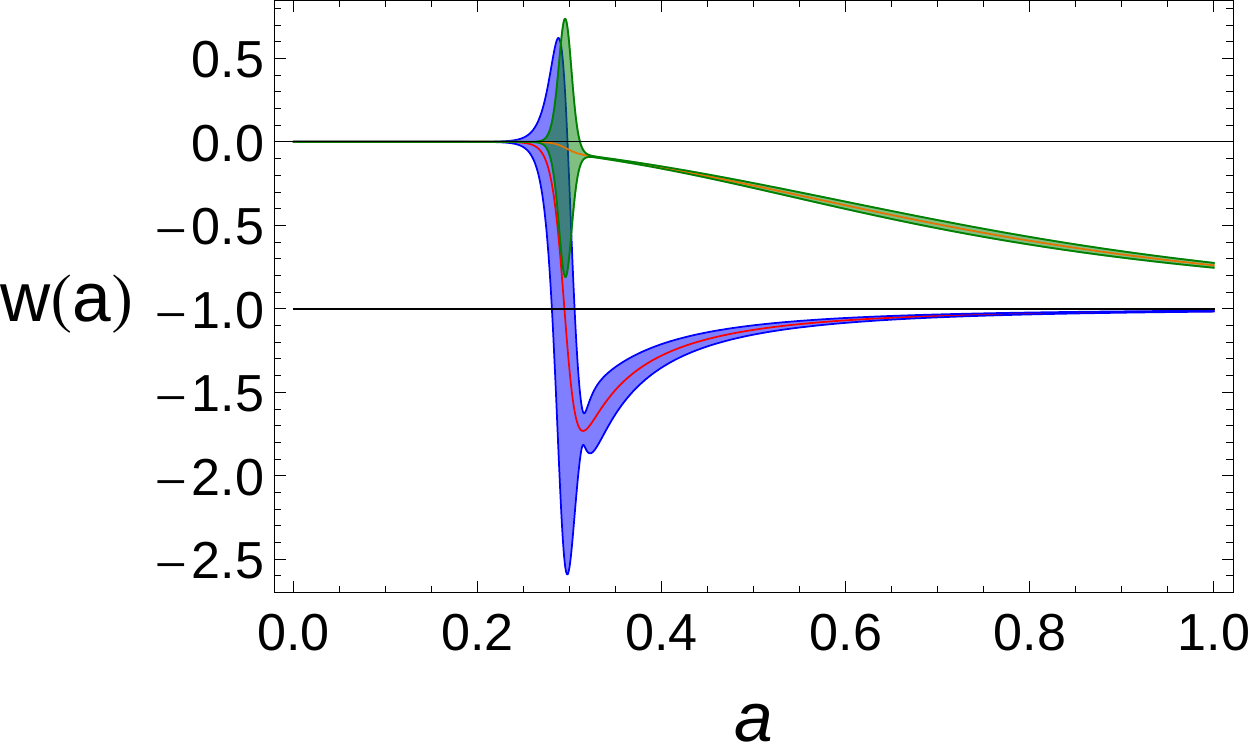}~~~~~
		\caption{Evolution of the EOS of the UDM fluid (green) for the best-fit model parameters, including derived uncertainty, and EOS of the DE section of the fluid (blue).} 
		\label{fig:w}
	\end{minipage}
\end{figure*}

We can also look at the dark energy sector of the UDM model. A DE density may be defined as in Eq.~(\ref{omegaDEdef}) and its value computed from Eq.~(\ref{omegaDE}) as a function of all the other parameters. The corresponding EOS is dynamical and can be computed from Eq.~(\ref{wDE}). The constraints on $\Omega_{\rm DE}$ and $w_{\rm DE}(a=1)$, derived from the MCMC chains, are shown in Table~\ref{table-UDM}. The evolution of $w_{\rm DE}$ for the best-fit parameter values is shown in Fig.~\ref{fig:w}, together with its 1-$\sigma$ variation. Notice that even though $w_{DE}$ is phantom after the fast transition, approaching $w_{DE} \sim -1$ today from the negative side, the UDM fluid does not violate the null energy condition because its EOS, also shown in Fig. \ref{fig:w}, does not cross the phantom divide.


We have thus a UDM model with fast transition that is viable given background data. Let us analyze now the behavior of its core parameters: the scale factor at the transition, $a_t$, and the rapidity of transition, $\beta$. Their constraints, also shown in Fig.~\ref{fig:UDM} and Table~\ref{table-UDM}, are weak. The 1-$\sigma$ interval for the transition redshift ranges from $z\sim 2$ to $z\sim 13$, while $\beta$ does not show a correlation with the other parameters. The posterior probability of $\beta$ shows a peaked structure. Looking in more detail into the likelihood values, we see the likelihood is essentially flat for $\beta > 1000$. The peaks in the $\beta$ posterior indicate the chain is not yet converged for this parameter, meaning there was not enough time to sample the unbound flat distribution and the chain remained occasionally stuck in some positions of the flat distribution. We have thus found that $\beta$ is unbound from above, which reflects the fact that for $\beta > \sim 1000$ the Hubble function is essentially identical for all $\beta$ values. On the other hand, $\beta$ is bound from below, we do not impose a $\beta > 0$ prior in the analysis.

\begin{figure}[h]
	\centering
	\begin{minipage}{0.85\textwidth}
		\centering
		\includegraphics[width=0.6\textwidth]{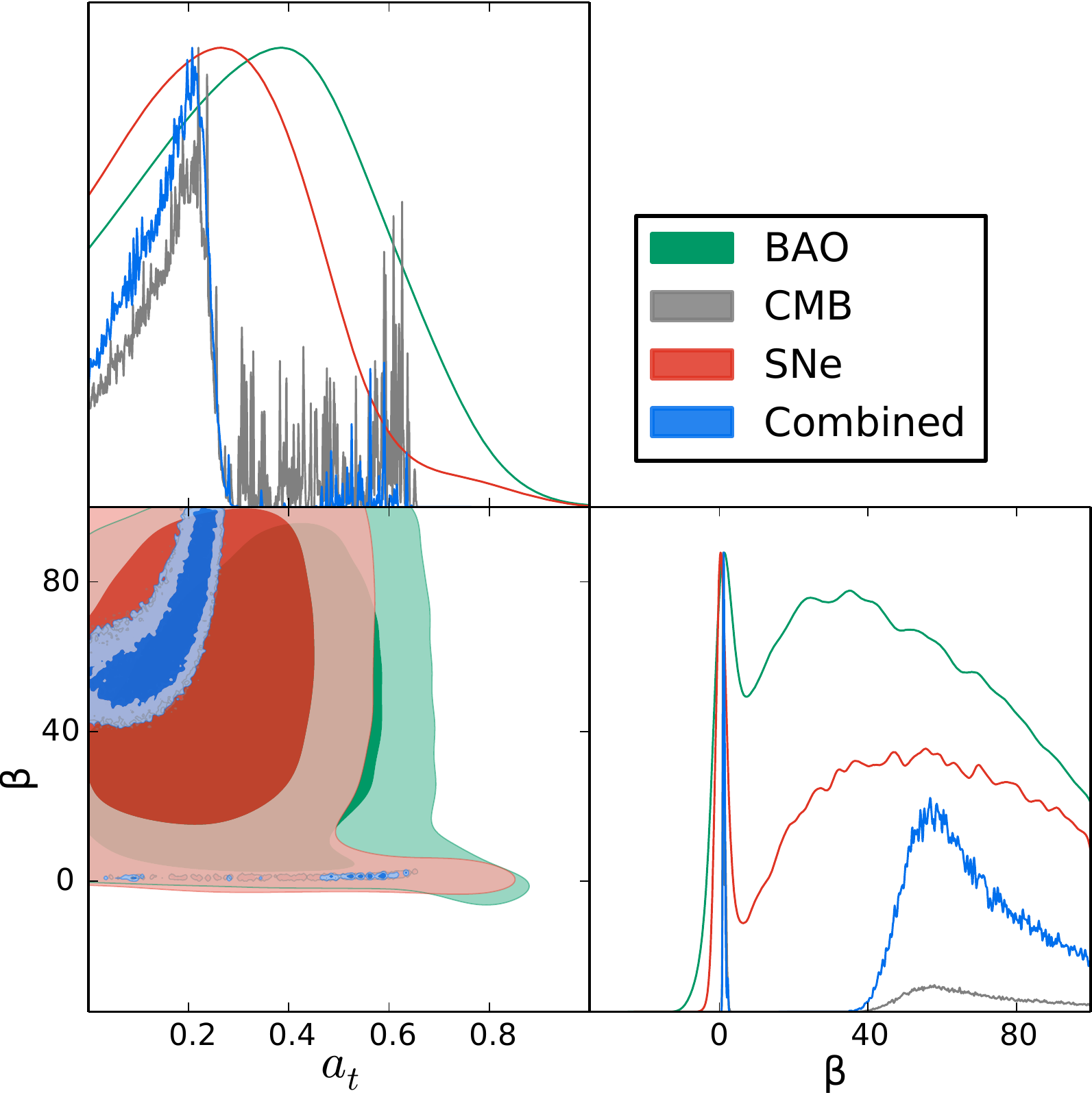}
		\caption{Posterior distribution from the slow transition MCMC chains.
Diagonal panels: One-dimensional marginalized posterior distributions for the UDM model parameters, for different data sets (SN, BAO, CMB, all combined). Off-diagonal panel: 1- and 2-$\sigma$ two-dimensional $(a_t,\beta)$ contours for the same data sets.} \label{fig:UDM-bgap}
	\end{minipage}
\end{figure}

These considerations led us to probe the low $\beta$ limit with better resolution. For this, we ran new chains considering only the range $\beta<100$. Given the low level of correlation with other parameters, we keep the density parameters fixed at the best-fit values, varying only $\beta$ and $a_t$. The scale factor at the transition must be kept free, since it is coupled with $\beta$ in the evolution of pressure and density, Eqs.~(\ref{ptanh}) and (\ref{tanh}), even though a degeneracy with $\beta$ does not show in Fig.~\ref{fig:UDM}. Notice also that this setup will artificially tighten the $a_t$ constraint due to its correlation with $\Omega_c$. We also ran separate chains for each data set and show the results in Fig.~\ref{fig:UDM-bgap}. We see now a sharp peak in the posterior of $\beta$ at $\beta=0$ that had not been picked up before. This point is basically a singularity in the space of UDM parameters. Indeed, in the $\beta=0$ limit, Eq.~(\ref{omegaUDM}) no longer presents a transition and the model reduces to $\Lambda$CDM, which explains its high likelihood. No transition, also means that the value of $a_t$ is meaningless, which explains the very narrow horizontal contour seen in the contour plot at $\beta=0$. As $\beta$ increases, the Hubble function starts to deviate from $\Lambda$CDM, until $\beta \sim 15$, and afterwards it approaches it again. This explains the dip in the $\beta$ posterior seen in all data sets. This effect is especially dramatic for the CMB shift parameters, which are able to reject the range $\beta < 40$.

Regarding $a_t$, the noisy structure seen in its posterior corresponds to the $\beta=0$ solution, while the rest of the probability volume lies along a well-defined degeneracy in the $(\beta, a_t)$ plane. Indeed, in this regime of low $\beta$ the data is able to pick up the degeneracy that arises from the fact that a slower transition needs to occur earlier in order to be able to reach today's density ratio. We fit the degeneracy direction with a cubic polynomial $\beta-\beta_0 = (a_t/0.22)^3$ to capture the $(\beta,a_t)$ dependence in the Hubble function, Eqs.~(\ref{friedman-2}) and (\ref{omegaUDM}). Here $\beta_0=54.6$ is the average chain value of $\beta$ for $a_t=0$, while $a_t=0.22$ is the median $a_t$ value quoted in Table~\ref{table-UDM}. With these assumptions, we find the following 1-$\sigma$ constraint:
\begin{equation}
(\beta-\beta_0)\,\left(\frac{a_t}{0.22}\right)^{-3} = 24.8 \pm 5.9\,.
\end{equation}

\begin{figure}[h]
	\centering
	\begin{minipage}{0.85\textwidth}
		\centering
		\includegraphics[width=0.6\textwidth]{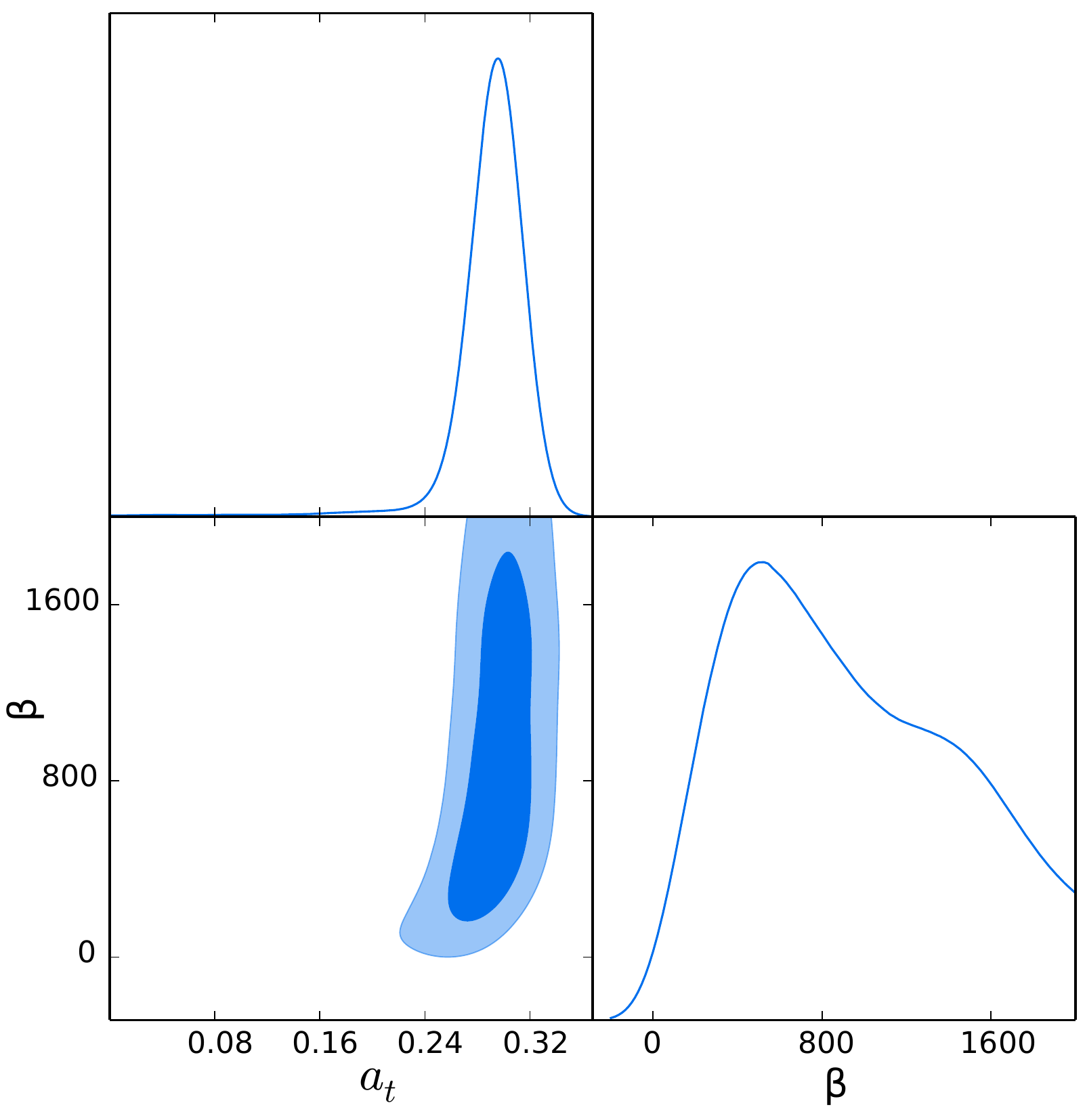}~~~~~
		\caption{Posterior distribution from the fast transition MCMC chains.
Diagonal panels: One-dimensional marginalized posterior distributions for the UDM model parameters. Off-diagonal panel: 1- and 2-$\sigma$ two-dimensional $(a_t,\beta)$ contours.} \label{fig:UDM-bmin}
	\end{minipage}
\end{figure}

We also need to look with higher resolution to the intermediate regime of $\beta$, to compare the likelihoods of the slow transition models with the fast transition ones. This is the regime of $\beta$ of a few hundreds, where the $\tanh$ function is not yet a step function. We thus ran a new $(a_t,\beta)$ chain restricted to $\beta < 2000$. The results of this analysis are shown in Fig.~\ref{fig:UDM-bmin}. The distribution of $a_t$ is now well constrained, showing a tight peak with a low-likelihood tail for low $a_t$ values. The tail corresponds to the slow transition regime studied in Fig.~\ref{fig:UDM-bgap}. This result then strongly favors intermediate and fast transitions over slow ones. This is confirmed by the posterior of $\beta$ that shows a strong increase from slow to fast transition, peaking around $\beta=600$. After the peak, the distribution falls down slowly with a long tail, which is just an effect of the strong prior $\beta < 2000$ imposed in this analysis, since the likelihood is essentially flat. We see then that the $\beta$ distribution is far from Gaussian and we can only find a lower limit for this parameter. From the $\Delta \chi^2$ values, we find a 1-$\sigma$ lower bound of $\beta > 300$.

\section{Conclusions}
\label{sec:concl}

In recent years, UDM models, for which DM and DE are described by a single dark fluid,  have become increasingly popular and drawn a considerable amount of attention. These models are undoubtedly promising candidates as effective theories. In this work, we have constrained a UDM scalar field model with a fast transition. The scalar field used has a noncanonical kinetic term in its Lagrangian and accounts for both the accelerated expansion of the Universe at late times and the clustering properties of the large-scale structure of the Universe at early times. The fast transition occurs between an Einstein-de Sitter CDM-like epoch and a late accelerated DE-like epoch and allows one to have a sufficiently small Jeans length, even if the speed of sound is large during the transition, because this happens so quickly that its effect is negligible.

In this study we investigated the regimes of slow and fast transition and assessed if they were distinguishable at background level. For this analysis we tested the models using supernovae Ia, baryon acoustic oscillations and CMB distance data. We have found a lower bound constraint for the rapidity of the transition $\beta > 300$, independent of the transition redshift. Slow transition models $\beta < 40$ were ruled out, while low-likelihood intermediate rapidity models featured a correlation between the transition redshift and rapidity.

The evidence of this model was compared to the evidence of $\Lambda$CDM and a phenomenological fast transition UDM model that was previously shown to be a good fit to background data. In both comparisons the model fared well, with no conclusive evidence against it.

The preference found for the fast transition regime, which is the condition required for enabling structure formation, together with the fact that the model has a similar evidence to $\Lambda$CDM and is a k-essence type physically motivated model with a well-defined Lagrangian, makes it an interesting and viable fundamental cosmological model.

For completeness, it is worth mentioning that k-essence-type models suffer in general from the development of caustics in the nonlinear regime \cite{Babichev:2016hys} (see however \cite{Mukohyama:2016ipl}). That is to say, characteristics of equations of motion cross at some finite time rendering the k-essence scalar field no longer single valued and consequently second derivatives of the field diverge. That would suggest that k-essence models cannot be considered as fundamental. However, this issue could possibly be solved by making the metric dynamical, such that gravitational backreaction would prevent the formation of caustics \cite{Babichev:2016hys}. Another possible way out of this problem was recently proposed in \cite{Babichev:2017lrx}, by introducing a complex scalar field such that the singularity does not develop in the real time and the real time evolution always remains smooth.

\acknowledgments
We thank Vincenzo Salzano for the use of his nested sampling code and Ruth Lazkoz and Diogo Castel\~ao for discussions. We also thank the anonymous referee for having raised the point of the inconsistency of the name 'unified dark matter'. 
This  work  was  supported  by  Funda\c{c}\~ao  para  a  Ci\^encia e a Tecnologia (FCT) through the research Grant No. UID/FIS/04434/2013. I.T. acknowledges support from FCT through the Investigador FCT Contract No. IF/01518/2014 and POPH/FSE (EC) by FEDER funding through the program COMPETE.
A.R.F. gratefully acknowledges support from FCT through Fellowship No. SFRH/BPD/96981/2103 (Portugal) and from Ministerio de Econom\'ia y Competitividad (Spain) through Project No. FIS2012-38816. 
I.L. acknowledges financial support through research Projects No. FIS2014-57956-P (comprising FEDER funds) from Ministerio de Econom\'ia y Competitividad and No. GIC17/116-IT956-16 from the Basque Government. I.L. further acknowledges financial support from the University of the Basque Country (UPV/EHU) through PhD Grant No. 750/2014, and from FCT through the Exploratory Project No. IF/01518/2014 (GLUE) during his stay at Instituto de Astrof\'isica e Ci\^encias do Espa\c{c}o, Faculdade de Ci\^encias da Universidade de Lisboa where part of this work was carried out.

This article is based upon work from COST Action CA15117 (CANTATA), supported by COST (European Cooperation in Science and Technology).

\bibliographystyle{apsrev}

\bibliography{UDM}

\begin{thebibliography}{65}
\expandafter\ifx\csname natexlab\endcsname\relax\def\natexlab#1{#1}\fi
\expandafter\ifx\csname bibnamefont\endcsname\relax
  \def\bibnamefont#1{#1}\fi
\expandafter\ifx\csname bibfnamefont\endcsname\relax
  \def\bibfnamefont#1{#1}\fi
\expandafter\ifx\csname citenamefont\endcsname\relax
  \def\citenamefont#1{#1}\fi
\expandafter\ifx\csname url\endcsname\relax
  \def\url#1{\texttt{#1}}\fi
\expandafter\ifx\csname urlprefix\endcsname\relax\def\urlprefix{URL }\fi
\providecommand{\bibinfo}[2]{#2}
\providecommand{\eprint}[2][]{\url{#2}}

\bibitem[{\citenamefont{{Weinberg} et~al.}(2013)\citenamefont{{Weinberg},
  {Mortonson}, {Eisenstein}, {Hirata}, {Riess}, and {Rozo}}}]{DWeinberg:2013}
\bibinfo{author}{\bibfnamefont{D.~H.} \bibnamefont{{Weinberg}}},
  \bibinfo{author}{\bibfnamefont{M.~J.} \bibnamefont{{Mortonson}}},
  \bibinfo{author}{\bibfnamefont{D.~J.} \bibnamefont{{Eisenstein}}},
  \bibinfo{author}{\bibfnamefont{C.}~\bibnamefont{{Hirata}}},
  \bibinfo{author}{\bibfnamefont{A.~G.} \bibnamefont{{Riess}}},
  \bibnamefont{and} \bibinfo{author}{\bibfnamefont{E.}~\bibnamefont{{Rozo}}},
  \bibinfo{journal}{Phys. Rep.} \textbf{\bibinfo{volume}{530}},
  \bibinfo{pages}{87} (\bibinfo{year}{2013}), \eprint{1201.2434}.

\bibitem[{\citenamefont{{Bull} et~al.}(2016)\citenamefont{{Bull}, {Akrami},
  {Adamek}, {Baker}, {Bellini}, {Beltr{\'a}n Jim{\'e}nez}, {Bentivegna},
  {Camera}, {Clesse}, {Davis} et~al.}}]{Bull:2016}
\bibinfo{author}{\bibfnamefont{P.}~\bibnamefont{{Bull}}},
  \bibinfo{author}{\bibfnamefont{Y.}~\bibnamefont{{Akrami}}},
  \bibinfo{author}{\bibfnamefont{J.}~\bibnamefont{{Adamek}}},
  \bibinfo{author}{\bibfnamefont{T.}~\bibnamefont{{Baker}}},
  \bibinfo{author}{\bibfnamefont{E.}~\bibnamefont{{Bellini}}},
  \bibinfo{author}{\bibfnamefont{J.}~\bibnamefont{{Beltr{\'a}n Jim{\'e}nez}}},
  \bibinfo{author}{\bibfnamefont{E.}~\bibnamefont{{Bentivegna}}},
  \bibinfo{author}{\bibfnamefont{S.}~\bibnamefont{{Camera}}},
  \bibinfo{author}{\bibfnamefont{S.}~\bibnamefont{{Clesse}}},
  \bibinfo{author}{\bibfnamefont{J.~H.} \bibnamefont{{Davis}}},
  \bibnamefont{et~al.}, \bibinfo{journal}{Physics of the Dark Universe}
  \textbf{\bibinfo{volume}{12}}, \bibinfo{pages}{56} (\bibinfo{year}{2016}),
  \eprint{1512.05356}.

\bibitem[{\citenamefont{Copeland et~al.}(2006)\citenamefont{Copeland, Sami, and
  Tsujikawa}}]{Copeland:2006wr}
\bibinfo{author}{\bibfnamefont{E.~J.} \bibnamefont{Copeland}},
  \bibinfo{author}{\bibfnamefont{M.}~\bibnamefont{Sami}}, \bibnamefont{and}
  \bibinfo{author}{\bibfnamefont{S.}~\bibnamefont{Tsujikawa}},
  \bibinfo{journal}{Int. J. Mod. Phys.} \textbf{\bibinfo{volume}{D15}},
  \bibinfo{pages}{1753} (\bibinfo{year}{2006}), \eprint{hep-th/0603057}.

\bibitem[{\citenamefont{Clifton et~al.}(2012)\citenamefont{Clifton, Ferreira,
  Padilla, and Skordis}}]{Clifton:2011jh}
\bibinfo{author}{\bibfnamefont{T.}~\bibnamefont{Clifton}},
  \bibinfo{author}{\bibfnamefont{P.~G.} \bibnamefont{Ferreira}},
  \bibinfo{author}{\bibfnamefont{A.}~\bibnamefont{Padilla}}, \bibnamefont{and}
  \bibinfo{author}{\bibfnamefont{C.}~\bibnamefont{Skordis}},
  \bibinfo{journal}{Phys. Rept.} \textbf{\bibinfo{volume}{513}},
  \bibinfo{pages}{1} (\bibinfo{year}{2012}), \eprint{1106.2476}.

\bibitem[{\citenamefont{Gonz\'{a}lez-D\'{i}az and
  Rozas-Fern\'{a}ndez}(2006)}]{GonzalezDiaz:2006tr}
\bibinfo{author}{\bibfnamefont{P.~F.} \bibnamefont{Gonz\'{a}lez-D\'{i}az}}
  \bibnamefont{and}
  \bibinfo{author}{\bibfnamefont{A.}~\bibnamefont{Rozas-Fern\'{a}ndez}},
  \bibinfo{journal}{Phys. Lett.} \textbf{\bibinfo{volume}{B641}},
  \bibinfo{pages}{134} (\bibinfo{year}{2006}), \eprint{astro-ph/0609263}.

\bibitem[{\citenamefont{Gonz\'{a}lez-D\'{i}az and
  Rozas-Fern\'{a}ndez}(2008)}]{GonzalezDiaz:2008ci}
\bibinfo{author}{\bibfnamefont{P.~F.} \bibnamefont{Gonz\'{a}lez-D\'{i}az}}
  \bibnamefont{and}
  \bibinfo{author}{\bibfnamefont{A.}~\bibnamefont{Rozas-Fern\'{a}ndez}},
  \bibinfo{journal}{Class. Quant. Grav.} \textbf{\bibinfo{volume}{25}},
  \bibinfo{pages}{175023} (\bibinfo{year}{2008}), \eprint{0807.2055}.

\bibitem[{\citenamefont{Gonz\'{a}lez-D\'{i}az and
  Rozas-Fern\'{a}ndez}(2014)}]{GonzalezDiaz:2008ba}
\bibinfo{author}{\bibfnamefont{P.~F.} \bibnamefont{Gonz\'{a}lez-D\'{i}az}}
  \bibnamefont{and}
  \bibinfo{author}{\bibfnamefont{A.}~\bibnamefont{Rozas-Fern\'{a}ndez}},
  \bibinfo{journal}{Phys. Lett.} \textbf{\bibinfo{volume}{B733}},
  \bibinfo{pages}{84} (\bibinfo{year}{2014}), \eprint{0811.2948}.

\bibitem[{\citenamefont{Rozas-Fern{\'a}ndez}(2017)}]{RozasFernandez2017}
\bibinfo{author}{\bibfnamefont{A.}~\bibnamefont{Rozas-Fern{\'a}ndez}},
  \bibinfo{journal}{General Relativity and Gravitation}
  \textbf{\bibinfo{volume}{49}}, \bibinfo{pages}{93} (\bibinfo{year}{2017}),
  ISSN \bibinfo{issn}{1572-9532},
  \urlprefix\url{http://dx.doi.org/10.1007/s10714-017-2254-3}.

\bibitem[{\citenamefont{{Smale} and {Wiltshire}}(2011)}]{Smale:2011}
\bibinfo{author}{\bibfnamefont{P.~R.} \bibnamefont{{Smale}}} \bibnamefont{and}
  \bibinfo{author}{\bibfnamefont{D.~L.} \bibnamefont{{Wiltshire}}},
  \bibinfo{journal}{Mon.Not.Roy.Astron.Soc.} \textbf{\bibinfo{volume}{413}},
  \bibinfo{pages}{367} (\bibinfo{year}{2011}), \eprint{1009.5855}.

\bibitem[{\citenamefont{Makler et~al.}(2003)\citenamefont{Makler, Quinet~de
  Oliveira, and Waga}}]{Makler:2002jv}
\bibinfo{author}{\bibfnamefont{M.}~\bibnamefont{Makler}},
  \bibinfo{author}{\bibfnamefont{S.}~\bibnamefont{Quinet~de Oliveira}},
  \bibnamefont{and} \bibinfo{author}{\bibfnamefont{I.}~\bibnamefont{Waga}},
  \bibinfo{journal}{Phys. Lett.} \textbf{\bibinfo{volume}{B555}},
  \bibinfo{pages}{1} (\bibinfo{year}{2003}), \eprint{astro-ph/0209486}.

\bibitem[{\citenamefont{Bertacca et~al.}(2010)\citenamefont{Bertacca, Bartolo,
  and Matarrese}}]{Bertacca:2010ct}
\bibinfo{author}{\bibfnamefont{D.}~\bibnamefont{Bertacca}},
  \bibinfo{author}{\bibfnamefont{N.}~\bibnamefont{Bartolo}}, \bibnamefont{and}
  \bibinfo{author}{\bibfnamefont{S.}~\bibnamefont{Matarrese}},
  \bibinfo{journal}{Adv. Astron.} \textbf{\bibinfo{volume}{2010}},
  \bibinfo{pages}{904379} (\bibinfo{year}{2010}), \eprint{1008.0614}.

\bibitem[{\citenamefont{Kamenshchik et~al.}(2001)\citenamefont{Kamenshchik,
  Moschella, and Pasquier}}]{Kamenshchik:2001cp}
\bibinfo{author}{\bibfnamefont{A.~{\relax Yu}.} \bibnamefont{Kamenshchik}},
  \bibinfo{author}{\bibfnamefont{U.}~\bibnamefont{Moschella}},
  \bibnamefont{and} \bibinfo{author}{\bibfnamefont{V.}~\bibnamefont{Pasquier}},
  \bibinfo{journal}{Phys. Lett.} \textbf{\bibinfo{volume}{B511}},
  \bibinfo{pages}{265} (\bibinfo{year}{2001}), \eprint{gr-qc/0103004}.

\bibitem[{\citenamefont{Bilic et~al.}(2002)\citenamefont{Bilic, Tupper, and
  Viollier}}]{Bilic:2001cg}
\bibinfo{author}{\bibfnamefont{N.}~\bibnamefont{Bilic}},
  \bibinfo{author}{\bibfnamefont{G.~B.} \bibnamefont{Tupper}},
  \bibnamefont{and} \bibinfo{author}{\bibfnamefont{R.~D.}
  \bibnamefont{Viollier}}, \bibinfo{journal}{Phys. Lett.}
  \textbf{\bibinfo{volume}{B535}}, \bibinfo{pages}{17} (\bibinfo{year}{2002}),
  \eprint{astro-ph/0111325}.

\bibitem[{\citenamefont{Bento et~al.}(2002)\citenamefont{Bento, Bertolami, and
  Sen}}]{Bento:2002ps}
\bibinfo{author}{\bibfnamefont{M.~C.} \bibnamefont{Bento}},
  \bibinfo{author}{\bibfnamefont{O.}~\bibnamefont{Bertolami}},
  \bibnamefont{and} \bibinfo{author}{\bibfnamefont{A.~A.} \bibnamefont{Sen}},
  \bibinfo{journal}{Phys. Rev.} \textbf{\bibinfo{volume}{D66}},
  \bibinfo{pages}{043507} (\bibinfo{year}{2002}), \eprint{gr-qc/0202064}.

\bibitem[{\citenamefont{Steinhardt}(1997)}]{cp}
\bibinfo{author}{\bibfnamefont{P.}~\bibnamefont{Steinhardt}},
  \emph{\bibinfo{title}{Critical Problems in Physics}}
  (\bibinfo{publisher}{Princeton University Press, Princeton.},
  \bibinfo{year}{1997}).

\bibitem[{\citenamefont{Camera et~al.}(2017)\citenamefont{Camera, Martinelli,
  and Bertacca}}]{Camera:2017tws}
\bibinfo{author}{\bibfnamefont{S.}~\bibnamefont{Camera}},
  \bibinfo{author}{\bibfnamefont{M.}~\bibnamefont{Martinelli}},
  \bibnamefont{and} \bibinfo{author}{\bibfnamefont{D.}~\bibnamefont{Bertacca}}
  (\bibinfo{year}{2017}), \eprint{1704.06277}.

\bibitem[{\citenamefont{Hu}(1998)}]{Hu:1998kj}
\bibinfo{author}{\bibfnamefont{W.}~\bibnamefont{Hu}},
  \bibinfo{journal}{Astrophys. J.} \textbf{\bibinfo{volume}{506}},
  \bibinfo{pages}{485} (\bibinfo{year}{1998}), \eprint{astro-ph/9801234}.

\bibitem[{\citenamefont{Garriga and Mukhanov}(1999)}]{Garriga:1999vw}
\bibinfo{author}{\bibfnamefont{J.}~\bibnamefont{Garriga}} \bibnamefont{and}
  \bibinfo{author}{\bibfnamefont{V.~F.} \bibnamefont{Mukhanov}},
  \bibinfo{journal}{Phys. Lett.} \textbf{\bibinfo{volume}{B458}},
  \bibinfo{pages}{219} (\bibinfo{year}{1999}), \eprint{hep-th/9904176}.

\bibitem[{\citenamefont{Pietrobon et~al.}(2008)\citenamefont{Pietrobon, Balbi,
  Bruni, and Quercellini}}]{Pietrobon:2008js}
\bibinfo{author}{\bibfnamefont{D.}~\bibnamefont{Pietrobon}},
  \bibinfo{author}{\bibfnamefont{A.}~\bibnamefont{Balbi}},
  \bibinfo{author}{\bibfnamefont{M.}~\bibnamefont{Bruni}}, \bibnamefont{and}
  \bibinfo{author}{\bibfnamefont{C.}~\bibnamefont{Quercellini}},
  \bibinfo{journal}{Phys. Rev.} \textbf{\bibinfo{volume}{D78}},
  \bibinfo{pages}{083510} (\bibinfo{year}{2008}), \eprint{0807.5077}.

\bibitem[{\citenamefont{Bertacca and Bartolo}(2007)}]{Bertacca:2007cv}
\bibinfo{author}{\bibfnamefont{D.}~\bibnamefont{Bertacca}} \bibnamefont{and}
  \bibinfo{author}{\bibfnamefont{N.}~\bibnamefont{Bartolo}},
  \bibinfo{journal}{JCAP} \textbf{\bibinfo{volume}{0711}}, \bibinfo{pages}{026}
  (\bibinfo{year}{2007}), \eprint{0707.4247}.

\bibitem[{\citenamefont{Carturan and Finelli}(2003)}]{Carturan:2002si}
\bibinfo{author}{\bibfnamefont{D.}~\bibnamefont{Carturan}} \bibnamefont{and}
  \bibinfo{author}{\bibfnamefont{F.}~\bibnamefont{Finelli}},
  \bibinfo{journal}{Phys. Rev.} \textbf{\bibinfo{volume}{D68}},
  \bibinfo{pages}{103501} (\bibinfo{year}{2003}), \eprint{astro-ph/0211626}.

\bibitem[{\citenamefont{Sandvik et~al.}(2004)\citenamefont{Sandvik, Tegmark,
  Zaldarriaga, and Waga}}]{Sandvik:2002jz}
\bibinfo{author}{\bibfnamefont{H.}~\bibnamefont{Sandvik}},
  \bibinfo{author}{\bibfnamefont{M.}~\bibnamefont{Tegmark}},
  \bibinfo{author}{\bibfnamefont{M.}~\bibnamefont{Zaldarriaga}},
  \bibnamefont{and} \bibinfo{author}{\bibfnamefont{I.}~\bibnamefont{Waga}},
  \bibinfo{journal}{Phys. Rev.} \textbf{\bibinfo{volume}{D69}},
  \bibinfo{pages}{123524} (\bibinfo{year}{2004}), \eprint{astro-ph/0212114}.

\bibitem[{\citenamefont{Scherrer}(2004)}]{Scherrer:2004au}
\bibinfo{author}{\bibfnamefont{R.~J.} \bibnamefont{Scherrer}},
  \bibinfo{journal}{Phys. Rev. Lett.} \textbf{\bibinfo{volume}{93}},
  \bibinfo{pages}{011301} (\bibinfo{year}{2004}), \eprint{astro-ph/0402316}.

\bibitem[{\citenamefont{Giannakis and Hu}(2005)}]{Giannakis:2005kr}
\bibinfo{author}{\bibfnamefont{D.}~\bibnamefont{Giannakis}} \bibnamefont{and}
  \bibinfo{author}{\bibfnamefont{W.}~\bibnamefont{Hu}}, \bibinfo{journal}{Phys.
  Rev.} \textbf{\bibinfo{volume}{D72}}, \bibinfo{pages}{063502}
  (\bibinfo{year}{2005}), \eprint{astro-ph/0501423}.

\bibitem[{\citenamefont{Piattella}(2010)}]{Piattella:2009da}
\bibinfo{author}{\bibfnamefont{O.~F.} \bibnamefont{Piattella}},
  \bibinfo{journal}{JCAP} \textbf{\bibinfo{volume}{1003}}, \bibinfo{pages}{012}
  (\bibinfo{year}{2010}), \eprint{0906.4430}.

\bibitem[{\citenamefont{Bertacca et~al.}(2008)\citenamefont{Bertacca, Bartolo,
  Diaferio, and Matarrese}}]{Bertacca:2008uf}
\bibinfo{author}{\bibfnamefont{D.}~\bibnamefont{Bertacca}},
  \bibinfo{author}{\bibfnamefont{N.}~\bibnamefont{Bartolo}},
  \bibinfo{author}{\bibfnamefont{A.}~\bibnamefont{Diaferio}}, \bibnamefont{and}
  \bibinfo{author}{\bibfnamefont{S.}~\bibnamefont{Matarrese}},
  \bibinfo{journal}{JCAP} \textbf{\bibinfo{volume}{0810}}, \bibinfo{pages}{023}
  (\bibinfo{year}{2008}), \eprint{0807.1020}.

\bibitem[{\citenamefont{Camera et~al.}(2009)\citenamefont{Camera, Bertacca,
  Diaferio, Bartolo, and Matarrese}}]{Camera:2009uz}
\bibinfo{author}{\bibfnamefont{S.}~\bibnamefont{Camera}},
  \bibinfo{author}{\bibfnamefont{D.}~\bibnamefont{Bertacca}},
  \bibinfo{author}{\bibfnamefont{A.}~\bibnamefont{Diaferio}},
  \bibinfo{author}{\bibfnamefont{N.}~\bibnamefont{Bartolo}}, \bibnamefont{and}
  \bibinfo{author}{\bibfnamefont{S.}~\bibnamefont{Matarrese}},
  \bibinfo{journal}{Mon. Not. Roy. Astron. Soc.}
  \textbf{\bibinfo{volume}{399}}, \bibinfo{pages}{1995} (\bibinfo{year}{2009}),
  \eprint{0902.4204}.

\bibitem[{\citenamefont{Camera et~al.}(2011)\citenamefont{Camera, Kitching,
  Heavens, Bertacca, and Diaferio}}]{Camera:2010wm}
\bibinfo{author}{\bibfnamefont{S.}~\bibnamefont{Camera}},
  \bibinfo{author}{\bibfnamefont{T.~D.} \bibnamefont{Kitching}},
  \bibinfo{author}{\bibfnamefont{A.~F.} \bibnamefont{Heavens}},
  \bibinfo{author}{\bibfnamefont{D.}~\bibnamefont{Bertacca}}, \bibnamefont{and}
  \bibinfo{author}{\bibfnamefont{A.}~\bibnamefont{Diaferio}},
  \bibinfo{journal}{Mon. Not. Roy. Astron. Soc.}
  \textbf{\bibinfo{volume}{415}}, \bibinfo{pages}{399} (\bibinfo{year}{2011}),
  \eprint{1002.4740}.

\bibitem[{\citenamefont{Radicella and Pavon}(2014)}]{Radicella:2014nka}
\bibinfo{author}{\bibfnamefont{N.}~\bibnamefont{Radicella}} \bibnamefont{and}
  \bibinfo{author}{\bibfnamefont{D.}~\bibnamefont{Pavon}},
  \bibinfo{journal}{Phys. Rev.} \textbf{\bibinfo{volume}{D89}},
  \bibinfo{pages}{067302} (\bibinfo{year}{2014}), \eprint{1403.2601}.

\bibitem[{\citenamefont{Piattella et~al.}(2010)\citenamefont{Piattella,
  Bertacca, Bruni, and Pietrobon}}]{Piattella:2009kt}
\bibinfo{author}{\bibfnamefont{O.~F.} \bibnamefont{Piattella}},
  \bibinfo{author}{\bibfnamefont{D.}~\bibnamefont{Bertacca}},
  \bibinfo{author}{\bibfnamefont{M.}~\bibnamefont{Bruni}}, \bibnamefont{and}
  \bibinfo{author}{\bibfnamefont{D.}~\bibnamefont{Pietrobon}},
  \bibinfo{journal}{JCAP} \textbf{\bibinfo{volume}{1001}}, \bibinfo{pages}{014}
  (\bibinfo{year}{2010}), \eprint{0911.2664}.

\bibitem[{\citenamefont{Bertacca et~al.}(2011)\citenamefont{Bertacca, Bruni,
  Piattella, and Pietrobon}}]{Bertacca:2010mt}
\bibinfo{author}{\bibfnamefont{D.}~\bibnamefont{Bertacca}},
  \bibinfo{author}{\bibfnamefont{M.}~\bibnamefont{Bruni}},
  \bibinfo{author}{\bibfnamefont{O.~F.} \bibnamefont{Piattella}},
  \bibnamefont{and}
  \bibinfo{author}{\bibfnamefont{D.}~\bibnamefont{Pietrobon}},
  \bibinfo{journal}{JCAP} \textbf{\bibinfo{volume}{1102}}, \bibinfo{pages}{018}
  (\bibinfo{year}{2011}), \eprint{1011.6669}.

\bibitem[{\citenamefont{Chiba et~al.}(2000)\citenamefont{Chiba, Okabe, and
  Yamaguchi}}]{Chiba:1999ka}
\bibinfo{author}{\bibfnamefont{T.}~\bibnamefont{Chiba}},
  \bibinfo{author}{\bibfnamefont{T.}~\bibnamefont{Okabe}}, \bibnamefont{and}
  \bibinfo{author}{\bibfnamefont{M.}~\bibnamefont{Yamaguchi}},
  \bibinfo{journal}{Phys. Rev.} \textbf{\bibinfo{volume}{D62}},
  \bibinfo{pages}{023511} (\bibinfo{year}{2000}), \eprint{astro-ph/9912463}.

\bibitem[{\citenamefont{Armendariz-Picon
  et~al.}(2001)\citenamefont{Armendariz-Picon, Mukhanov, and
  Steinhardt}}]{ArmendarizPicon:2000ah}
\bibinfo{author}{\bibfnamefont{C.}~\bibnamefont{Armendariz-Picon}},
  \bibinfo{author}{\bibfnamefont{V.~F.} \bibnamefont{Mukhanov}},
  \bibnamefont{and} \bibinfo{author}{\bibfnamefont{P.~J.}
  \bibnamefont{Steinhardt}}, \bibinfo{journal}{Phys. Rev.}
  \textbf{\bibinfo{volume}{D63}}, \bibinfo{pages}{103510}
  (\bibinfo{year}{2001}), \eprint{astro-ph/0006373}.

\bibitem[{\citenamefont{Kang et~al.}(2007)\citenamefont{Kang, Vanchurin, and
  Winitzki}}]{Kang:2007vs}
\bibinfo{author}{\bibfnamefont{J.~U.} \bibnamefont{Kang}},
  \bibinfo{author}{\bibfnamefont{V.}~\bibnamefont{Vanchurin}},
  \bibnamefont{and} \bibinfo{author}{\bibfnamefont{S.}~\bibnamefont{Winitzki}},
  \bibinfo{journal}{Phys. Rev.} \textbf{\bibinfo{volume}{D76}},
  \bibinfo{pages}{083511} (\bibinfo{year}{2007}), \eprint{0706.3994}.

\bibitem[{\citenamefont{Cruz et~al.}(2009)\citenamefont{Cruz,
  Gonz\'{a}lez-D\'{i}az, Rozas-Fern\'{a}ndez, and Sanchez}}]{Cruz:2008cwa}
\bibinfo{author}{\bibfnamefont{N.}~\bibnamefont{Cruz}},
  \bibinfo{author}{\bibfnamefont{P.~F.} \bibnamefont{Gonz\'{a}lez-D\'{i}az}},
  \bibinfo{author}{\bibfnamefont{A.}~\bibnamefont{Rozas-Fern\'{a}ndez}},
  \bibnamefont{and} \bibinfo{author}{\bibfnamefont{G.}~\bibnamefont{Sanchez}},
  \bibinfo{journal}{Phys. Lett.} \textbf{\bibinfo{volume}{B679}},
  \bibinfo{pages}{293} (\bibinfo{year}{2009}), \eprint{0812.4856}.

\bibitem[{\citenamefont{Rozas-Fern\'{a}ndez}(2012)}]{RozasFernandez:2011je}
\bibinfo{author}{\bibfnamefont{A.}~\bibnamefont{Rozas-Fern\'{a}ndez}},
  \bibinfo{journal}{Phys. Lett.} \textbf{\bibinfo{volume}{B709}},
  \bibinfo{pages}{313} (\bibinfo{year}{2012}), \eprint{1106.0056}.

\bibitem[{\citenamefont{Rozas-Fern\'{a}ndez}(2014)}]{Rozas-Fernandez:2014tsa}
\bibinfo{author}{\bibfnamefont{A.}~\bibnamefont{Rozas-Fern\'{a}ndez}},
  \bibinfo{journal}{Gen. Rel. Grav.} \textbf{\bibinfo{volume}{46}},
  \bibinfo{pages}{1825} (\bibinfo{year}{2014}), \eprint{1410.6373}.

\bibitem[{\citenamefont{Diez-Tejedor and Feinstein}(2005)}]{DiezTejedor:2005fz}
\bibinfo{author}{\bibfnamefont{A.}~\bibnamefont{Diez-Tejedor}}
  \bibnamefont{and}
  \bibinfo{author}{\bibfnamefont{A.}~\bibnamefont{Feinstein}},
  \bibinfo{journal}{Int. J. Mod. Phys.} \textbf{\bibinfo{volume}{D14}},
  \bibinfo{pages}{1561} (\bibinfo{year}{2005}), \eprint{gr-qc/0501101}.

\bibitem[{\citenamefont{Bilic}(2008)}]{Bilic:2008zk}
\bibinfo{author}{\bibfnamefont{N.}~\bibnamefont{Bilic}},
  \bibinfo{journal}{Phys. Rev.} \textbf{\bibinfo{volume}{D78}},
  \bibinfo{pages}{105012} (\bibinfo{year}{2008}), \eprint{0806.0642}.

\bibitem[{\citenamefont{Bruni et~al.}(2013)\citenamefont{Bruni, Lazkoz, and
  Rozas-Fernandez}}]{Bruni:2012sn}
\bibinfo{author}{\bibfnamefont{M.}~\bibnamefont{Bruni}},
  \bibinfo{author}{\bibfnamefont{R.}~\bibnamefont{Lazkoz}}, \bibnamefont{and}
  \bibinfo{author}{\bibfnamefont{A.}~\bibnamefont{Rozas-Fernandez}},
  \bibinfo{journal}{Mon. Not. Roy. Astron. Soc.}
  \textbf{\bibinfo{volume}{431}}, \bibinfo{pages}{2907} (\bibinfo{year}{2013}),
  \eprint{1210.1880}.

\bibitem[{\citenamefont{Liddle and Urena-Lopez}(2006)}]{Liddle:2006qz}
\bibinfo{author}{\bibfnamefont{A.~R.} \bibnamefont{Liddle}} \bibnamefont{and}
  \bibinfo{author}{\bibfnamefont{L.~A.} \bibnamefont{Urena-Lopez}},
  \bibinfo{journal}{Phys. Rev. Lett.} \textbf{\bibinfo{volume}{97}},
  \bibinfo{pages}{161301} (\bibinfo{year}{2006}), \eprint{astro-ph/0605205}.

\bibitem[{\citenamefont{{Amendola} et~al.}(2016)\citenamefont{{Amendola},
  {Appleby}, {Avgoustidis}, {Bacon}, {Baker}, {Baldi}, {Bartolo}, {Blanchard},
  {Bonvin}, {Borgani} et~al.}}]{Amendola:2016saw}
\bibinfo{author}{\bibfnamefont{L.}~\bibnamefont{{Amendola}}},
  \bibinfo{author}{\bibfnamefont{S.}~\bibnamefont{{Appleby}}},
  \bibinfo{author}{\bibfnamefont{A.}~\bibnamefont{{Avgoustidis}}},
  \bibinfo{author}{\bibfnamefont{D.}~\bibnamefont{{Bacon}}},
  \bibinfo{author}{\bibfnamefont{T.}~\bibnamefont{{Baker}}},
  \bibinfo{author}{\bibfnamefont{M.}~\bibnamefont{{Baldi}}},
  \bibinfo{author}{\bibfnamefont{N.}~\bibnamefont{{Bartolo}}},
  \bibinfo{author}{\bibfnamefont{A.}~\bibnamefont{{Blanchard}}},
  \bibinfo{author}{\bibfnamefont{C.}~\bibnamefont{{Bonvin}}},
  \bibinfo{author}{\bibfnamefont{S.}~\bibnamefont{{Borgani}}},
  \bibnamefont{et~al.}, \bibinfo{journal}{ArXiv e-prints}
  (\bibinfo{year}{2016}), \eprint{1606.00180}.

\bibitem[{\citenamefont{{Lazkoz} et~al.}(2016)\citenamefont{{Lazkoz},
  {Leanizbarrutia}, and {Salzano}}}]{Lazkoz:2016hmh}
\bibinfo{author}{\bibfnamefont{R.}~\bibnamefont{{Lazkoz}}},
  \bibinfo{author}{\bibfnamefont{I.}~\bibnamefont{{Leanizbarrutia}}},
  \bibnamefont{and}
  \bibinfo{author}{\bibfnamefont{V.}~\bibnamefont{{Salzano}}},
  \bibinfo{journal}{Phys. Rev.} \textbf{\bibinfo{volume}{D93}},
  \bibinfo{eid}{043537} (\bibinfo{year}{2016}), \eprint{1602.01331}.

\bibitem[{\citenamefont{Christensen et~al.}(2001)\citenamefont{Christensen,
  Meyer, Knox, and Luey}}]{Christensen:2001gj}
\bibinfo{author}{\bibfnamefont{N.}~\bibnamefont{Christensen}},
  \bibinfo{author}{\bibfnamefont{R.}~\bibnamefont{Meyer}},
  \bibinfo{author}{\bibfnamefont{L.}~\bibnamefont{Knox}}, \bibnamefont{and}
  \bibinfo{author}{\bibfnamefont{B.}~\bibnamefont{Luey}},
  \bibinfo{journal}{Class.Quant.Grav.} \textbf{\bibinfo{volume}{18}},
  \bibinfo{pages}{2677} (\bibinfo{year}{2001}), \eprint{astro-ph/0103134}.

\bibitem[{\citenamefont{Lewis and Bridle}(2002)}]{Lewis:2002ah}
\bibinfo{author}{\bibfnamefont{A.}~\bibnamefont{Lewis}} \bibnamefont{and}
  \bibinfo{author}{\bibfnamefont{S.}~\bibnamefont{Bridle}},
  \bibinfo{journal}{Phys.Rev.} \textbf{\bibinfo{volume}{D66}},
  \bibinfo{pages}{103511} (\bibinfo{year}{2002}), \eprint{astro-ph/0205436}.

\bibitem[{\citenamefont{Suzuki et~al.}(2012)}]{Suzuki:2011hu}
\bibinfo{author}{\bibfnamefont{N.}~\bibnamefont{Suzuki}} \bibnamefont{et~al.},
  \bibinfo{journal}{Astrophys. J.} \textbf{\bibinfo{volume}{746}},
  \bibinfo{pages}{85} (\bibinfo{year}{2012}), \eprint{1105.3470}.

\bibitem[{\citenamefont{Conley et~al.}(2011)}]{Conley:2011ku}
\bibinfo{author}{\bibfnamefont{A.}~\bibnamefont{Conley}} \bibnamefont{et~al.}
  (\bibinfo{collaboration}{SNLS Collaboration}),
  \bibinfo{journal}{Astrophys.J.Suppl.} \textbf{\bibinfo{volume}{192}},
  \bibinfo{pages}{1} (\bibinfo{year}{2011}), \eprint{1104.1443}.

\bibitem[{\citenamefont{Blake et~al.}(2012)}]{Blake:2012pj}
\bibinfo{author}{\bibfnamefont{C.}~\bibnamefont{Blake}} \bibnamefont{et~al.},
  \bibinfo{journal}{Mon. Not. Roy. Astron. Soc.}
  \textbf{\bibinfo{volume}{425}}, \bibinfo{pages}{405} (\bibinfo{year}{2012}),
  \eprint{1204.3674}.

\bibitem[{\citenamefont{Wang and Wang}(2013)}]{Wang2013}
\bibinfo{author}{\bibfnamefont{Y.}~\bibnamefont{Wang}} \bibnamefont{and}
  \bibinfo{author}{\bibfnamefont{S.}~\bibnamefont{Wang}},
  \bibinfo{journal}{Phys.Rev.} \textbf{\bibinfo{volume}{D88}},
  \bibinfo{pages}{043522} (\bibinfo{year}{2013}), \eprint{1304.4514}.

\bibitem[{\citenamefont{{Wang} and {Mukherjee}}(2007)}]{WangMukherjee2007}
\bibinfo{author}{\bibfnamefont{Y.}~\bibnamefont{{Wang}}} \bibnamefont{and}
  \bibinfo{author}{\bibfnamefont{P.}~\bibnamefont{{Mukherjee}}},
  \bibinfo{journal}{Phys. Rev.} \textbf{\bibinfo{volume}{D76}},
  \bibinfo{eid}{103533} (\bibinfo{year}{2007}), \eprint{astro-ph/0703780}.

\bibitem[{\citenamefont{Hu and Sugiyama}(1996)}]{Hu:1995en}
\bibinfo{author}{\bibfnamefont{W.}~\bibnamefont{Hu}} \bibnamefont{and}
  \bibinfo{author}{\bibfnamefont{N.}~\bibnamefont{Sugiyama}},
  \bibinfo{journal}{Astrophys.J.} \textbf{\bibinfo{volume}{471}},
  \bibinfo{pages}{542} (\bibinfo{year}{1996}), \eprint{astro-ph/9510117}.

\bibitem[{\citenamefont{Fixsen}(2009)}]{Fixsen:2009ug}
\bibinfo{author}{\bibfnamefont{D.}~\bibnamefont{Fixsen}},
  \bibinfo{journal}{Astrophys.J.} \textbf{\bibinfo{volume}{707}},
  \bibinfo{pages}{916} (\bibinfo{year}{2009}), \eprint{0911.1955}.

\bibitem[{\citenamefont{Ade et~al.}(2014)}]{Ade:2013zuv}
\bibinfo{author}{\bibfnamefont{P.}~\bibnamefont{Ade}} \bibnamefont{et~al.}
  (\bibinfo{collaboration}{Planck Collaboration}),
  \bibinfo{journal}{Astron.Astrophys.} \textbf{\bibinfo{volume}{571}},
  \bibinfo{pages}{A16} (\bibinfo{year}{2014}), \eprint{1303.5076}.

\bibitem[{\citenamefont{{Komatsu} et~al.}(2011)\citenamefont{{Komatsu},
  {Smith}, {Dunkley}, {Bennett}, {Gold}, {Hinshaw}, {Jarosik}, {Larson},
  {Nolta}, {Page} et~al.}}]{wmap7}
\bibinfo{author}{\bibfnamefont{E.}~\bibnamefont{{Komatsu}}},
  \bibinfo{author}{\bibfnamefont{K.~M.} \bibnamefont{{Smith}}},
  \bibinfo{author}{\bibfnamefont{J.}~\bibnamefont{{Dunkley}}},
  \bibinfo{author}{\bibfnamefont{C.~L.} \bibnamefont{{Bennett}}},
  \bibinfo{author}{\bibfnamefont{B.}~\bibnamefont{{Gold}}},
  \bibinfo{author}{\bibfnamefont{G.}~\bibnamefont{{Hinshaw}}},
  \bibinfo{author}{\bibfnamefont{N.}~\bibnamefont{{Jarosik}}},
  \bibinfo{author}{\bibfnamefont{D.}~\bibnamefont{{Larson}}},
  \bibinfo{author}{\bibfnamefont{M.~R.} \bibnamefont{{Nolta}}},
  \bibinfo{author}{\bibfnamefont{L.}~\bibnamefont{{Page}}},
  \bibnamefont{et~al.}, \bibinfo{journal}{Astrophys.J.Suppl.}
  \textbf{\bibinfo{volume}{192}}, \bibinfo{eid}{18} (\bibinfo{year}{2011}),
  \eprint{1001.4538}.

\bibitem[{\citenamefont{{Bennett} et~al.}(2013)\citenamefont{{Bennett},
  {Larson}, {Weiland}, {Jarosik}, {Hinshaw}, {Odegard}, {Smith}, {Hill},
  {Gold}, {Halpern} et~al.}}]{wmap9}
\bibinfo{author}{\bibfnamefont{C.~L.} \bibnamefont{{Bennett}}},
  \bibinfo{author}{\bibfnamefont{D.}~\bibnamefont{{Larson}}},
  \bibinfo{author}{\bibfnamefont{J.~L.} \bibnamefont{{Weiland}}},
  \bibinfo{author}{\bibfnamefont{N.}~\bibnamefont{{Jarosik}}},
  \bibinfo{author}{\bibfnamefont{G.}~\bibnamefont{{Hinshaw}}},
  \bibinfo{author}{\bibfnamefont{N.}~\bibnamefont{{Odegard}}},
  \bibinfo{author}{\bibfnamefont{K.~M.} \bibnamefont{{Smith}}},
  \bibinfo{author}{\bibfnamefont{R.~S.} \bibnamefont{{Hill}}},
  \bibinfo{author}{\bibfnamefont{B.}~\bibnamefont{{Gold}}},
  \bibinfo{author}{\bibfnamefont{M.}~\bibnamefont{{Halpern}}},
  \bibnamefont{et~al.}, \bibinfo{journal}{Astrophys.J.Suppl.}
  \textbf{\bibinfo{volume}{208}}, \bibinfo{eid}{20} (\bibinfo{year}{2013}),
  \eprint{1212.5225}.

\bibitem[{\citenamefont{{Dunkley} et~al.}(2005)\citenamefont{{Dunkley},
  {Bucher}, {Ferreira}, {Moodley}, and {Skordis}}}]{Dunkley:2005}
\bibinfo{author}{\bibfnamefont{J.}~\bibnamefont{{Dunkley}}},
  \bibinfo{author}{\bibfnamefont{M.}~\bibnamefont{{Bucher}}},
  \bibinfo{author}{\bibfnamefont{P.~G.} \bibnamefont{{Ferreira}}},
  \bibinfo{author}{\bibfnamefont{K.}~\bibnamefont{{Moodley}}},
  \bibnamefont{and}
  \bibinfo{author}{\bibfnamefont{C.}~\bibnamefont{{Skordis}}},
  \bibinfo{journal}{Mon. Not. Roy. Astron. Soc.}
  \textbf{\bibinfo{volume}{356}}, \bibinfo{pages}{925} (\bibinfo{year}{2005}),
  \eprint{astro-ph/0405462}.

\bibitem[{\citenamefont{{Tereno} et~al.}(2005)\citenamefont{{Tereno},
  {Dor{\'e}}, {van Waerbeke}, and {Mellier}}}]{Tereno:2005}
\bibinfo{author}{\bibfnamefont{I.}~\bibnamefont{{Tereno}}},
  \bibinfo{author}{\bibfnamefont{O.}~\bibnamefont{{Dor{\'e}}}},
  \bibinfo{author}{\bibfnamefont{L.}~\bibnamefont{{van Waerbeke}}},
  \bibnamefont{and}
  \bibinfo{author}{\bibfnamefont{Y.}~\bibnamefont{{Mellier}}},
  \bibinfo{journal}{Astronomy \& Astrophysics} \textbf{\bibinfo{volume}{429}},
  \bibinfo{pages}{383} (\bibinfo{year}{2005}), \eprint{astro-ph/0404317}.

\bibitem[{\citenamefont{Trotta}(2007)}]{Trotta:2005ar}
\bibinfo{author}{\bibfnamefont{R.}~\bibnamefont{Trotta}},
  \bibinfo{journal}{Mon.Not.Roy.Astron.Soc.} \textbf{\bibinfo{volume}{378}},
  \bibinfo{pages}{72} (\bibinfo{year}{2007}), \eprint{astro-ph/0504022}.

\bibitem[{\citenamefont{Mukherjee et~al.}(2006)\citenamefont{Mukherjee,
  Parkinson, and Liddle}}]{Mukherjee:2005wg}
\bibinfo{author}{\bibfnamefont{P.}~\bibnamefont{Mukherjee}},
  \bibinfo{author}{\bibfnamefont{D.}~\bibnamefont{Parkinson}},
  \bibnamefont{and} \bibinfo{author}{\bibfnamefont{A.~R.}
  \bibnamefont{Liddle}}, \bibinfo{journal}{Astrophys.J.}
  \textbf{\bibinfo{volume}{638}}, \bibinfo{pages}{L51} (\bibinfo{year}{2006}),
  \eprint{astro-ph/0508461}.

\bibitem[{\citenamefont{Gordon and Trotta}(2007)}]{Gordon:2007xm}
\bibinfo{author}{\bibfnamefont{C.}~\bibnamefont{Gordon}} \bibnamefont{and}
  \bibinfo{author}{\bibfnamefont{R.}~\bibnamefont{Trotta}},
  \bibinfo{journal}{Mon.Not.Roy.Astron.Soc.} \textbf{\bibinfo{volume}{382}},
  \bibinfo{pages}{1859} (\bibinfo{year}{2007}), \eprint{0706.3014}.

\bibitem[{\citenamefont{{Liddle}}(2004)}]{Liddle:2004}
\bibinfo{author}{\bibfnamefont{A.~R.} \bibnamefont{{Liddle}}},
  \bibinfo{journal}{Mon. Not. Roy. Astron. Soc.}
  \textbf{\bibinfo{volume}{351}}, \bibinfo{pages}{L49} (\bibinfo{year}{2004}),
  \eprint{astro-ph/0401198}.

\bibitem[{\citenamefont{{S{\'a}ez-G{\'o}mez}
  et~al.}(2016)\citenamefont{{S{\'a}ez-G{\'o}mez}, {Carvalho}, {Lobo}, and
  {Tereno}}}]{SaezGomez:2016}
\bibinfo{author}{\bibfnamefont{D.}~\bibnamefont{{S{\'a}ez-G{\'o}mez}}},
  \bibinfo{author}{\bibfnamefont{C.~S.} \bibnamefont{{Carvalho}}},
  \bibinfo{author}{\bibfnamefont{F.~S.~N.} \bibnamefont{{Lobo}}},
  \bibnamefont{and} \bibinfo{author}{\bibfnamefont{I.}~\bibnamefont{{Tereno}}},
  \bibinfo{journal}{Phys. Rev.} \textbf{\bibinfo{volume}{D94}},
  \bibinfo{eid}{024034} (\bibinfo{year}{2016}), \eprint{1603.09670}.

\bibitem[{\citenamefont{Babichev}(2016)}]{Babichev:2016hys}
\bibinfo{author}{\bibfnamefont{E.}~\bibnamefont{Babichev}},
  \bibinfo{journal}{JHEP} \textbf{\bibinfo{volume}{04}}, \bibinfo{pages}{129}
  (\bibinfo{year}{2016}), \eprint{1602.00735}.

\bibitem[{\citenamefont{Mukohyama et~al.}(2016)\citenamefont{Mukohyama, Namba,
  and Watanabe}}]{Mukohyama:2016ipl}
\bibinfo{author}{\bibfnamefont{S.}~\bibnamefont{Mukohyama}},
  \bibinfo{author}{\bibfnamefont{R.}~\bibnamefont{Namba}}, \bibnamefont{and}
  \bibinfo{author}{\bibfnamefont{Y.}~\bibnamefont{Watanabe}},
  \bibinfo{journal}{Phys. Rev.} \textbf{\bibinfo{volume}{D94}},
  \bibinfo{pages}{023514} (\bibinfo{year}{2016}), \eprint{1605.06418}.

\bibitem[{\citenamefont{{Babichev} and {Ramazanov}}(2017)}]{Babichev:2017lrx}
\bibinfo{author}{\bibfnamefont{E.}~\bibnamefont{{Babichev}}} \bibnamefont{and}
  \bibinfo{author}{\bibfnamefont{S.}~\bibnamefont{{Ramazanov}}},
  \bibinfo{journal}{ArXiv e-prints}  (\bibinfo{year}{2017}),
  \eprint{1704.03367}.

\end{thebibliography}

\end{document}